%% file: Dutson2012_astro-ph.tex
\documentclass[useAMS,usenatbib]{mn2e}

\newcommand{\fermi}{{\it Fermi }}
\newcommand{\fermilat}{{\it Fermi}-LAT }
\newcommand{\gam}{$\gamma$}
\newcommand\T{\rule{0pt}{3.1ex}}

\usepackage{longtable}
\usepackage{multicol}
\usepackage{booktabs}
\usepackage{multirow}
\usepackage{setspace}
\usepackage{epstopdf}
\usepackage[]{graphicx}
\usepackage{booktabs}
\usepackage{amssymb}
\usepackage{subfigure}

\title[A Stacked Analysis of BCGs Observed with the
{\rm Fermi}-LAT]{A Stacked Analysis of Brightest Cluster Galaxies
  Observed with the \textbfit{Fermi} Large Area Telescope}

\author[K. L. Dutson, R. J. White, A. C. Edge, J. A. Hinton and M. T. Hogan]{K. L. Dutson$^{1}$\thanks{E-mail: kate.dutson@leicester.ac.uk}, R. J. White$^{1}$, A. C. Edge$^{2}$, J. A. Hinton$^{1}$ and M. T. Hogan$^{2}$\\
$^{1}$Department of Physics and Astronomy, The University of Leicester, University Road, Leicester, LE1 7RH, UK\\
$^{2}$Department of Physics, Durham University, South Road, Durham, DH1 3LE, UK}
\begin{document}

\date{Accepted 2012 November 22. Received 2012 November 22; in original form 2012 September 14}

\pagerange{\pageref{firstpage}--\pageref{lastpage}} \pubyear{2012}

\maketitle

\label{firstpage}

\begin{abstract}
We present the results of a search for high-energy \gam-ray emission
from a large sample of galaxy clusters sharing the properties of three
existing \fermilat detections (in Perseus, Virgo and Abell\,3392),
namely a powerful radio source within their brightest cluster galaxy
(BCG). From a parent, X-ray flux-limited sample of 863 clusters, we
select 114 systems with a core-dominated BCG radio flux above
50/75~mJy (in the NVSS and SUMSS surveys, respectively), stacking data
from the first 45 months of the \fermi mission in three energy bands,
to determine statistical limits on the \gam-ray fluxes of the ensemble
of candidate sources.

For a $>$300 MeV selection, the distribution of detection significance
across the sample is consistent with that across control samples for
significances $<$3$\sigma$, but has a tail extending to higher values,
including three $>$4$\sigma$ signals which are not associated with
previously identified \gam-ray emission. Modelling of the data in
these fields results in the detection of four non-2FGL \fermi sources,
though none of these appear to be {\it unambiguously} associated with
the BCG candidate. Only one is sufficiently close to be a plausible
counterpart (RXC\,J0132.6$-$0804), and the remaining three appear to
be background AGN. A search at energies $>$3~GeV hints at emission
from the BCG in A\,2055, which hosts a BL~Lac object.

There is no evidence for a signal in the stacked data, and the upper
limit derived on the \gam-ray flux of an average radio-bright BCG in
each band is at least an order-of-magnitude more constraining than
that calculated for individual objects. $F_{1 \rm GeV}/F_{1.4 \rm
GHz}$ for an average BCG in the sample is $<$15, compared with
$\approx$120 for NGC\,1275 in Perseus, which might indicate a special
case for those objects detected at high energies. The tentative
suggestion that point-like {\it beamed} emission from member galaxies
comprise the dominant bright \gam-ray sources in clusters, implies
searches for evidence of dark matter annihilation or large scale
merger shock signatures for example, need to account for a significant
level of contamination from within each cluster, that is both highly
stochastic and varies significantly over time.
\end{abstract}

\begin{keywords}
galaxies: active -- galaxies: clusters: general -- gamma rays:
observations -- radio continuum: general -- radiation
mechanisms: non-thermal -- (galaxies:) cooling flows
\end{keywords}

\section{Introduction}

Feedback from the central active galactic nucleus (hereafter AGN)
within a cooling-core cluster of galaxies is considered to play an
integral large-scale role in the suppression of the observed cooling
flow \citep[see e.g.][]{fabian02,sijacki07,bower06}. The hot, diffuse,
intracluster gas, which comprises 90\% of the baryonic matter of the
cluster and is bound gravitationally by a cold dark matter (DM) halo,
is revealed observationally through thermal Bremsstrahlung emission in
the X-ray band. From velocity dispersion measurements, the atmosphere
of this intracluster medium (ICM) typically has a temperature between
$10^{7}$ and $10^{8}\,{\rm K}$ \citep{mcnamara07}. The gas loses
energy to radiation primarily in the X-ray band, condensing onto the
central galaxy, and sub-sonic, pressure-driven inflows develop within
the dense cooling region, where the radiative cooling time for such
emission $t_{{\rm cool}}$, is less than the age of the system
\citep{fabian94}.

However, as evidenced by spectroscopic data from
current-generation X-ray observatories {\it Chandra} and {\it
XMM-Newton}, emission predicted by the standard cooling-flow model
from expected quantities of cooling gas is not detected
\citep{peterson03}. This soft X-ray deficit necessitates some
reheating mechanism that acts to replenish the energy being radiated
away, thus quenching the cooling flow. An attractive agent for this
feedback is the central AGN, whose outbursts would not only resolve
the dearth of X-ray luminosity, but also account for the measured
truncation at the high-mass end of the galaxy luminosity function
\citep{benson03}.

Support for this explanation is given by high-resolution observations
of cavities in the X-ray halo of many systems
\citep[e.g.][]{fabian00,mcnamara00}, coincident with radio synchrotron
emission associated with the activity of the central engine. The AGN
in the brightest cluster galaxy (hereafter BCG) injects energy into
the ICM in the form of bubbles containing relativistic particles,
which displace the thermal gas. So-called `ghost cavities' have also
been found using {\it Chandra}, interpreted as relics of past AGN
activity; bubbles that have previously detached from the radio lobes
and risen buoyantly through the ICM, no longer bright at $1.4\,{\rm
GHz}$. They are now known to be filled with {\it low-frequency} radio
synchrotron emission \citep[e.g.][]{clarke05,wise07}. This evidence
implicates AGN outbursts as intermittent phenomena; the power
output averaged over time of the central engine able to balance the
radiative cooling of the cluster core. For a review of AGN heating in
clusters of galaxies, see \cite{mcnamara07}.

Both models of galaxy formation/clustering, and of feedback, and
observations (including those aforementioned in the hard X-ray and
radio bands), establish galaxy clusters as hosts to significant
non-thermal particle populations \citep{volk99,blasi08}. The remaining
tracer of cosmic ray (CR) acceleration is \gam~radiation, and
several emission scenarios predict BCGs to be sources of high-energy
(HE) and very-high-energy (VHE) photons, with spectra dependent on the
dominant energy content of the AGN jets; the nature of the seed
particles they inject into the ICM, the strength of the local
magnetic field, and the target density around the inflated bubbles
\citep[e.g.][]{hinton07}.

Ultra-relativistic particles may play an important role on several
different scales in and around BCGs in cooling-core clusters, for
example in AGN jets, mini-halos, and cluster-scale halos. \gam-ray
emission may arise in each via a number of dissimilar processes
\citep[e.g.][]{pfrommer04}. Primary CR electrons accelerated in jets,
or re-accelerated fossil CRs in mini-halos will produce inverse
Compton (IC) emission. Inelastic proton-proton collisions of {\it
hadronic} CRs on the target medium will lead to $\pi^{0}$ decay
emission and IC emission from secondary (or cascade)
\citep[e.g.][]{blasi99, aharonian02} electrons. Most of these
BCG-driven processes will produce GeV emission that is point-like for
current detectors. However, it should be noted that -- being massive
and DM-dominated -- clusters of galaxies are expected sources of {\it
extended} \gam-ray emission via DM annihilation
\citep{ackermann10a}. Similarly, treating galaxy clusters as long-term
reservoirs of CR hadrons, diffuse HE emission from $\pi^{0}$ decay of
particles accelerated (for example) in cluster-scale accretion shocks
is theorised \citep[e.g.][]{pfrommer04}. The various emission
scenarios produce different spectral, temporal and spatial signatures
at GeV energies. The repository of HE data afforded by the {\it Fermi
Gamma-ray Space Telescope} (hereafter {\it Fermi}) since its launch in
2008, is therefore an attractive resource when used in conjunction
with multiwavelength data. Whilst a detection of extended cluster
emission at GeV energies is yet to be made \citep{ackermann10b},
point-like emission from a number of BCGs is seen and there is
considerable potential for improving our understanding of this class
of objects using GeV data.

To date, excluding systems whose AGN jets are favourably aligned close
to the observer's line of sight so as to boost emission via
relativistic beaming -- associated with `BL Lac' objects within the
{\it unified model} of active galaxies \citep{antonucci93} -- only two
BCGs have been confirmed as emitters of HE \gam~rays.  These are
NGC\,1275 and M\,87 \citep{abdo09a,abdo09b}: the dominant member
galaxies of the cooling-core clusters Perseus and Virgo,
respectively. Both are known to host bright central radio sources
associated with the active nucleus (see \citet{vermeulen94,asada06}
and \citet{biretta91,sparks96}), and are also detected at very-high
energies \citep{aleksic12,aharonian06}. The multiwavelength emission
of AGN is characteristically variable, and indeed the \gam-ray flux is
found to vary on timescales down to days in the case of NGC\,1275
\citep{brown11}; similarly for M\,87
\citep{abramowski12}. Light-crossing-time arguments thus restrict the
physical size of the emitting region responsible for, at least a
component of, the \gam-ray emission.

These sources are HE laboratories for the class of BCGs, and motivate
a search for further instances of \gam-ray activity within it. The
purpose of this work is to conduct such a search, using data from the
Large Area Telescope (LAT) on board the \fermi satellite, accumulated
thus far over the mission lifetime.  To this end, we set out to
imitate a deeper observation than is currently achievable in HE
\gam~rays given the sensitivity of the {\it Fermi}-LAT, by
constructing a sample of many potential sources within the BCG class,
and stacking the output of individual analyses.

We defend our derivation of a suitable sample of candidate sources in
\S\ref{sec:sample}, and the \fermilat analysis procedure then
carried out for each target is described in \S\ref{sec:analysis}. The
results are presented in \S\ref{sec:results}, and discussed in
\S\ref{sec:discussion}. Conclusions of the work are then drawn in
\S\ref{sec:conclusion}.

\section[]{Target Selection}
\label{sec:sample}

Considering NGC\,1275 to be prototypical of \gam-ray sources
within the class of BCGs, our selection criteria were therefore
designed to identify similar objects. The key requirements were the
presence of a massive cluster of galaxies and evidence of non-thermal
activity in the BCG.  We constructed a sample of
candidates drawn from five parent catalogues of X-ray flux-limited
galaxy clusters: the Brightest 55 (B55) Sample \citep{edge90}, the
ROSAT Bright Cluster Sample (BCS) \citep{ebeling98}, the extended
ROSAT Bright Cluster Sample (eBCS) \citep{ebeling00}, the ROSAT-ESO
Flux-Limited X-ray (REFLEX) Galaxy Cluster Survey Catalogue
\citep{bohringer04}, and the MAssive Cluster Survey (MACS)
\citep{ebeling01}, encompassing objects from both the Southern and
Northern hemispheres and including the brightest clusters from low
Galactic latitudes. We include two clusters (RXC\,J1350.3+0940 and
RXC\,J1832.5+6848) that meet the eBCS flux limit but were mis-identified
as AGN when the sample was published, due to the presence of a bright
radio source, and for comparison, two X-ray-luminous clusters that
contain known AGN (4C+55.16 \citep{hlavacek11}, and E\,1821+643
\citep{russell10}) though the cluster emission falls just below the                  
eBCS limit once the AGN contribution is taken into account.

From this sample of 863 unique clusters (accounting for the small
overlap between these five samples), we selected the 151 systems in
which the BCG is detected above $50\,{\rm mJy}$ in the NRAO VLA Sky
Survey (NVSS) \citep{condon98} or $75\,{\rm mJy}$ in the Sydney
University Molonglo Sky Survey (SUMSS) \citep{mauch03}. We took care
to remove systems where the radio emission is from a cluster member
projected close to the BCG. Of this sample we have higher resolution
radio imaging from the VLA or ATCA telescopes at 5 and/or $8\,{\rm
GHz}$ for the majority (114 of 151), which can be used to identify
those sources with a significant, flat spectrum core component
\citep{hogan}.

We have optical spectra for 148 of the 151 BCGs recognised by our
radio selection, and of these 64\% (97 objects), exhibit optical line
emission. This is significantly higher than the 25--30\% of BCGs in an
X-ray selected sample presenting line emission, indicating that our
radio selection strongly prefers clusters with a cooling flow
\citep{burns90,cavagnolo08,mittal09,hogan}.

\begin{figure*}
  \centering
  \includegraphics[width=17cm]{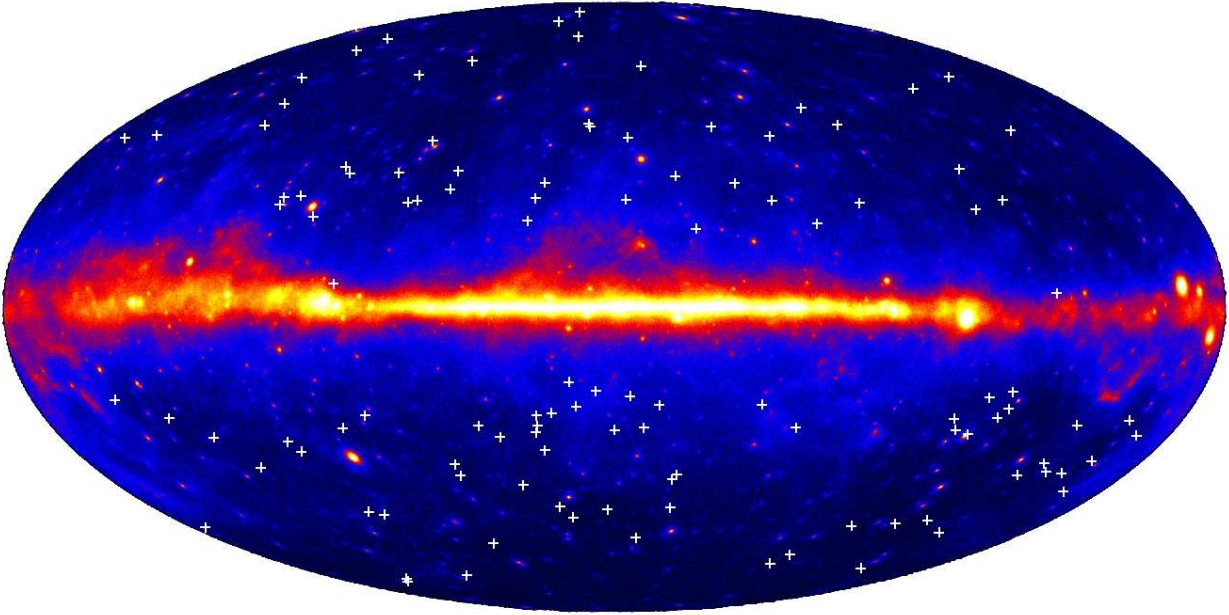}
  \caption{An all-sky map constructed from $\sim$45 months of
  $300\,{\rm MeV} \leq E \leq 300\,{\rm GeV}$ \fermi data, scaled
  logarithmically. The BCG positions of the 114 candidates of the main
  sample are indicated by the white crosses.}
  \label{fig:allsky}
\end{figure*}

The sample represents a complete selection of radio-bright clusters
that is dominated by cooling-core clusters, spanning a wide range in
redshift ($0.009 < z < 0.45$). It can be divided into BCGs whose radio
emission has some contribution from the core or is core-dominated, and
those with extended radio morphology and a lobe-dominated radio
flux. The former comprise 114 objects and constitute the `main sample'
(see Figure~\ref{fig:allsky} and Table~\ref{tab:sample}), with the
remaining 37 BCGs forming an `extended sample'. We compiled a counter
sample of 65 `control' clusters drawn from the parent X-ray sample
that do {\it not} contain a radio-bright BCG, nor optical emission
lines, but share the same redshift distribution as the radio bright
sample for a representative comparison. We also generated a pair of
control samples of 114 candidate sources each with random positions on
the sky (generated by performing first a positive and then a negative
5\degr~shift in Galactic latitude on the coordinates of the main
sample candidates) for direct comparison with the main sample.

The parent clusters of the main sample BCGs include hosts to five
BL~Lac objects, two of which are known HE \gam-ray sources listed in
the \fermilat Second Source Catalog (2FGL) \citep{nolan12}:
2FGL\,J0627.1$-$3528 in A\,3392 and 2FGL\,J1958.4-3012 in
RXC\,J1958.2$-$3011. The other three BL~Lacs, RGB\,J1144+676 in A1366,
RGB\,J1518+062 in A2055 and B2\,2334+23 in A\,2627, are not known
\fermi sources. These BL~Lacs are included as the original cluster
identification from the ROSAT Survey incorrectly attributed the (bulk
of the) X-ray emission to the cluster
\citep[e.g. RXC\,J1958.2$-$3011,][]{bohringer04}. However, in all five
cases, pointed {\it Chandra} or {\it XMM-Newton} observations indicate
that the BL~Lac dominates the emission, so the cluster does not meet
the X-ray flux limit applied to select the other clusters. Therefore
even this small \fermilat detection fraction is an upper limit, given
only NGC\,1275 (a misaligned blazar) and M\,87 (a radio galaxy) are
dominated by the extended X-ray emission from their host cluster, and
hence meet the cluster selection criteria.  For the purpose of the
analyses detailed hereafter, these two well-studied BCGs are not
included in the sample, since our aim is to identify whether any
additional sources in the class are detected using \fermi. There is no
{\it clear} cluster around RXC\,J1958.2$-$3011, so its classification
as a BCG is not certain, but with this caveat in mind it is kept,
particularly as an additional \fermi source to check the output of
analyses against.

\section[]{\textbfit {F\lowercase{ermi}}-LAT Observations and Data Analysis}
\label{sec:analysis}

The \fermilat is a pair conversion instrument sensitive to photons of
energy $\sim$20 MeV -- $>$300 GeV, with a wide field of view (2.4
sr) and large effective area ($\approx$8000 ${\rm cm^{2}}$ on axis at
$1\,{\rm GeV}$). The detector is described in detail in
\citet{atwood09} and references therein. \fermi observations are taken
primarily in {\it all-sky-survey} mode, in which the spacecraft is
pointed at some angle from the zenith, and rocked north and south of
its orbital plane, scanning the whole sky every $\sim$3 hours.

The 151 BCGs of our main and extended samples were treated as
candidate point sources of HE \gam~rays, and each analysed using the
\fermi Science Tools version v9r27p1, and instrument response function
(IRF) P7SOURCE$\_$V6. Event and spacecraft {\it all-sky-survey} data
amassed between the 4$^{\rm th}$ of August 2008 and the 28$^{\rm th}$ of
April 2012 (a Mission-Elapsed-Time interval of 239557417 to 357308125,
equating to $\sim$1362 days) were extracted for a source region (SR)
of 12\degr, centred on each BCG position (taken from optical data in
the literature) and in the energy range $300\,{\rm MeV} \leq E \leq
300\,{\rm GeV}$ (thus discarding events at the lowest energies, where
the high point spread function (PSF) results in considerable source
confusion). A zenith cut of 100\degr~was applied to eliminate
\gam~rays from the Earth's limb, and all `Source' class events were
considered.

For each candidate source the data were fitted to a source model
constructed to describe (in addition to potential emission from the
target BCG), the emission from nearby (i.e. within the SR) 2FGL
sources, and the diffuse Galactic and extragalactic (isotropic)
\gam-ray backgrounds (thus recent models gal\_2yearp7v6\_v0.fits and
iso\_p7v6source.txt
respectively\footnote{http://fermi.gsfc.nasa.gov/ssc/data/access/lat/BackgroundModels.html}
were utilised; their normalisation free to vary). The photon index and
normalisation for objects within 1\degr~of the candidate position
were allowed to vary. Outside of this region but within the specified
{\it region of interest} (ROI) of 8\degr, the normalisation was free
to vary, but the photon index was fixed to the 2FGL value. Sources
outside of the ROI but within the SR were included in the model, but
their aforementioned parameters were fixed to the catalogue values. In
the event that a candidate position coincided with a 2FGL source
(within 0.2\degr), then that source was automatically removed from the
model. This action provides a useful check of the output of our
analyses against the catalogue values for known \fermi sources, whilst
allowing all BCGs in the sample to be treated in the same way.

The maximum-likelihood spectral estimator {\sc gtlike} was used to
perform a {\it binned} fit, modelling the candidate source emission
with a power law spectrum. {\sc gtmodel} was then used to obtain a
model map of each candidate region given the result of the fit. To
construct counts maps from the data {\sc gtbin} was utilised. Binned
exposure cubes were also generated (through use of the {\sc
gtexpcube2} tool) as part of the binned likelihood analysis chain. A
binned method of likelihood fitting was adopted over an unbinned one,
because it was found to be more robust, the fit converging
irrespective of how many 2FGL sources a given SR contained.

The distribution of Test Statistic (TS) values (output by {\sc
gtlike}) across the sample could then be studied. A critical TS of 25
(corresponding roughly to a detection significance of 5$\sigma$
\citep{mattox96}) was decided upon, below which an upper limit on the
source emission was automatically calculated. All candidate sources
with TS $\gtrapprox$16 were nevertheless investigated on an individual
basis: TS maps were constructed and any peaks found thereon fitted with
parabolic ellipsoids to localise the emission. Where appropriate the
data were then remodelled to constrain potential new {\it offset}
sources.

The ensemble of counts maps constructed from the sample data, and that
of the model maps output from fitting based on a {\it null} hypothesis
(that is, the absence of a candidate \fermi source at the position of
the BCG), were stacked. These stacked maps could then be sliced to
provide 1-D comparisons between the summed model fit and the data. To
derive an upper limit on the \gam-ray flux of an average source in our
sample using this stacking method, the stacked counts map was used to
obtain a number of {\it on-source} counts ($N_{{\rm on}}$) within a
region of radius comparable with the PSF of the LAT instrument on
average across the selected energy band. An estimate of the number of
background photons in this region, $\hat{N}_{{\rm B}}$ was obtained by
similarly integrating the stacked model map. The number of {\it
  off-source} counts ($N_{{\rm off}}$) is then $\hat{N}_{{\rm B}} /
\alpha$ where $alpha$ is taken as the ratio of the solid angle in the
chosen radius to that of the field of view. Thus an estimate of the
\gam-ray signal present is given by the excess counts, $N_{{\rm on}} -
\alpha \cdot N_{{\rm off}}$. A 95\% confidence upper limit on the
signal was computed following the method of \citet{rolke05}, assuming
only Poisson errors on the background estimate. Whilst a more
sophisticated likelihood method similar to that used by
\citet{huber12} may provide a modest sensitivity improvement, our
approach is simple and robust.

Weighted exposure maps were constructed for each candidate source and
stacked, to provide a summed exposure map from which the exposure
averaged over solid angle in the on-region could be determined. This
value was used to transform our signal upper limit to an upper limit
on the \gam-ray flux. The rms variation in exposure across the sample
is $\approx$6\%.

For the purpose of investigating potential weak, hard spectrum,
candidate source emission; less contaminated by the diffuse
background, and given that stacking is most promising in the
signal-limited regime, where every additional field adds potential
source photons with minimal cost in background contamination, energy
cuts above $3\,{\rm GeV}$ and $30\,{\rm GeV}$ were applied to the data
and the analyses repeated in an otherwise identical manner; upper
limits calculated within regions of radii comparable to the LAT PSF at
the respective lower energy bounds. Corresponding analyses were
performed simultaneously for the control and radio-quiet samples.

\section[]{Results}
\label{sec:results}

A summary of the source properties of the BCGs of the main sample;
their respective TS values and/or upper limits output from the
analysis procedure outlined in \S\ref{sec:analysis} is given in
Table~\ref{tab:sample}. For the sake of clarity, the test statistic is
shown as zero for sources with TS $<$1. An upper limit is included for
sources with TS $<$25. Those BCGs with TS $>$25 are in:

\begin{description}
\item A\,3392 
\item RXC\,J1958.2$-$3011
\item MACS\,J1347.5$-$1145
\end{description}

As noted in \S\ref{sec:sample}, the sources in A\,3392 and
RXC\,J1958.2$-$3011 correspond to the known \fermi sources
2FGL\,J0627.1$-$3528 and 2FGL\,J1958.4$-$3012,
respectively. The BCG in MACS\,J1347.5$-$1145 is {\it not} in the 2FGL
catalogue. These sources, and those that fall shy of the TS cut, but
for which TS is greater than $\approx$16 are discussed in
\S\ref{ssec:sources} below.

\begin{onecolumn}
\include{table}

\end{onecolumn}

\begin{twocolumn}

\subsection{Distribution of Detection Significance}
\label{ssec:dist}

The distribution of $\sqrt{{\rm TS}}$ (corresponding approximately to
the statistical significance, see \S\ref{sec:analysis}) for each
sample analysed is shown in Figure~\ref{fig:1}. The extended and
radio-quiet cluster samples have been scaled appropriately, and the
control samples averaged, so that comparisons may easily be made with
the main sample of 114 objects. Common to all distributions is a clear
peak at 0, and also a tail containing several signals at up to
$\approx$3$\sigma$. However this tail is more pronounced for the main
sample -- implying that a stacking analysis may be worthwhile (see
\S\ref{ssec:stacking} below) -- and outliers are present: the high-TS
sources are visible above 4$\sigma$ (see \S\ref{ssec:sources} below),
including the two \fermi sources shown in red; all members of the main
sample. The only exception is the extended sample BCG in S\,753, for which
a TS of 19.9 was derived. Inspection of the field containing this
object reveals a highly confused region of diffuse emission, likely
associated with the Galactic background, and as such this result is
not believed.
 
\begin{figure}
  \centering
  \includegraphics[width=7.5cm]{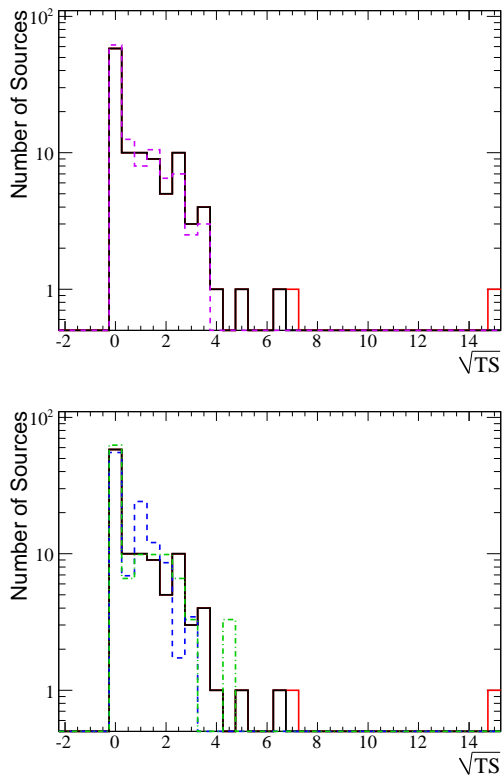}
  \caption{Histograms showing the comparison between the distribution
    of detection significance across the samples described in
    \S\ref{sec:sample}. The main sample of BCGs (black line) is
    included on both plots. {\bf Top Panel:} The (averaged) control
    sample of random positions on the sky (dashed violet line). {\bf
      Bottom Panel:} The scaled extended cluster sample (dot-dashed
    green line) and radio-quiet cluster sample (dashed blue line). The
    (two) sample BCGs with existing \fermi detections have been added
    in red.}
  \label{fig:1}
\end{figure}

\subsection{Stacking Analysis}
\label{ssec:stacking}

The stacked analyses of the main sample above the three energy cuts --
described in \S\ref{sec:analysis} -- are illustrated in
Figure~\ref{fig:2}. In each case the left panel shows the stacked
counts map in the colour scale, overlayed with contours defining the
corresponding stacked model map for the null hypothesis. The maps to
be summed were each centred on the candidate source position (0,0),
which is marked on the final stacked plot. The right panels show
one-dimensional slices through both the summed counts and model maps
(the area covered by these slices is indicated by the boxes drawn on
the stacked maps; their width motivated in each case by the LAT PSF as
described in \S\ref{sec:analysis}, to match the regions within which
upper limits are computed: 1\degr, 0.4\degr~and 0.2\degr~for the
$\geq$300~MeV, $\geq$3~GeV and $\geq$30~GeV datasets,
respectively). Sources with a high individual TS ($>$16), and those
within 0.2\degr~of a catalogue source are not included in the
stack. The field containing the BCG A\,3360 has also been removed, due
to the proximity (within 1\degr) of the very bright ($(3.71 \pm
0.07)\times10^{-8} {\rm ph\,cm}^{-2}{\rm s}^{-1}$) 2FGL source
2FGL\,J0538.8$-$4405. The model provides a reasonable description of
the data, and there is no evidence for an excess close to the
origin. Applying the method described in \S\ref{sec:analysis}, the
derived 95\% upper limits on the \gam-ray flux of an average
source within the class of BCGs, given the cuts applied and within a
region of size comparable with the LAT PSF, are found to be 7.4$\times
10^{-11} {\rm ph\,cm}^{-2}{\rm s}^{-1}$ (within a 1\degr~radius for
$300\,{\rm MeV} \leq E \leq 300\,{\rm GeV}$), 7.1$\times 10^{-12} {\rm
  ph\,cm}^{-2}{\rm s}^{-1}$ (within a 0.4\degr~radius for $3\,{\rm
  GeV} \leq E \leq 300\,{\rm GeV}$) and 3.7$\times 10^{-13} {\rm
  ph\,cm}^{-2} {\rm s}^{-1}$ (within a 0.2\degr~radius for $30\,{\rm
  GeV} \leq E \leq 300\,{\rm GeV}$).

\begin{figure*}    
  \begin{center}
    \subfigure{%
     \includegraphics[height=6.8cm]{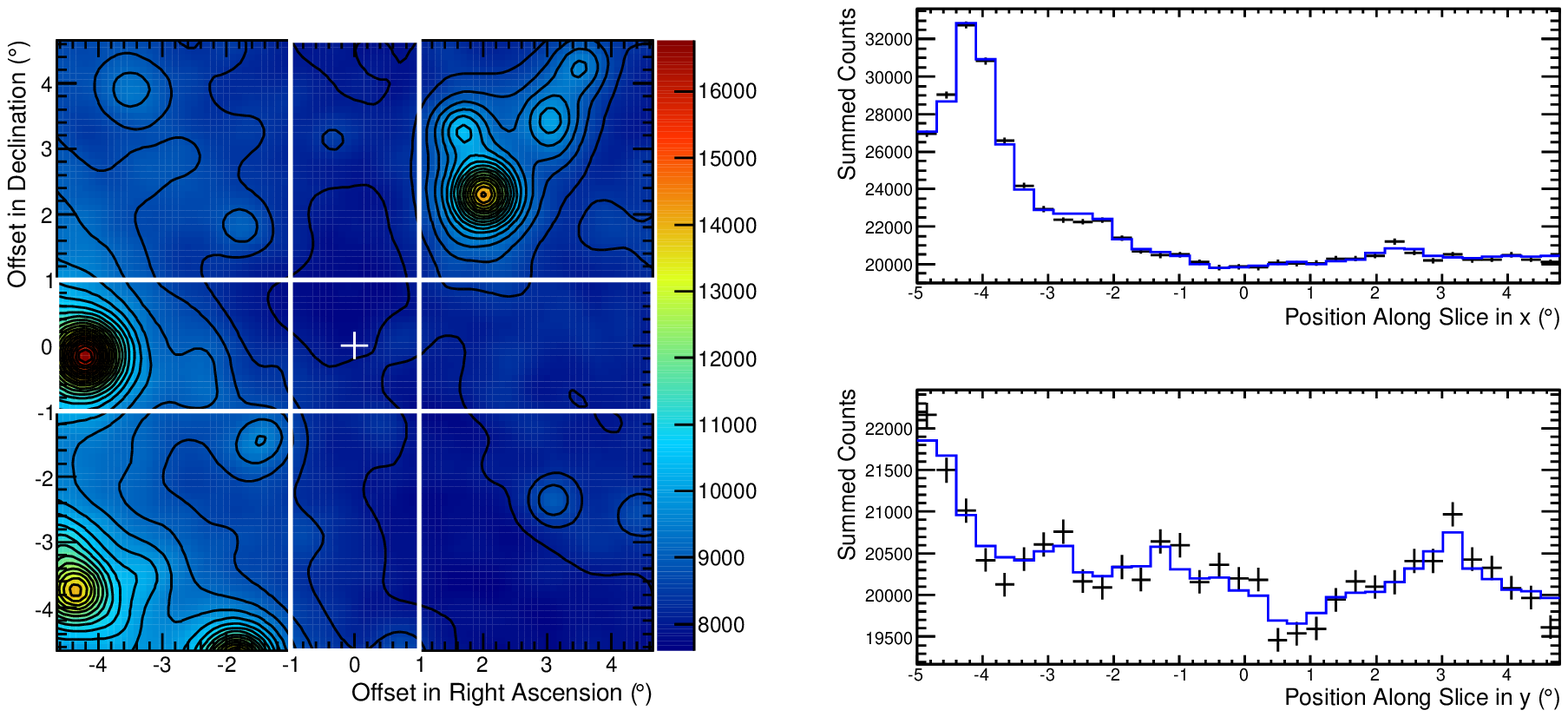}}
    \subfigure{%
     \includegraphics[height=6.8cm]{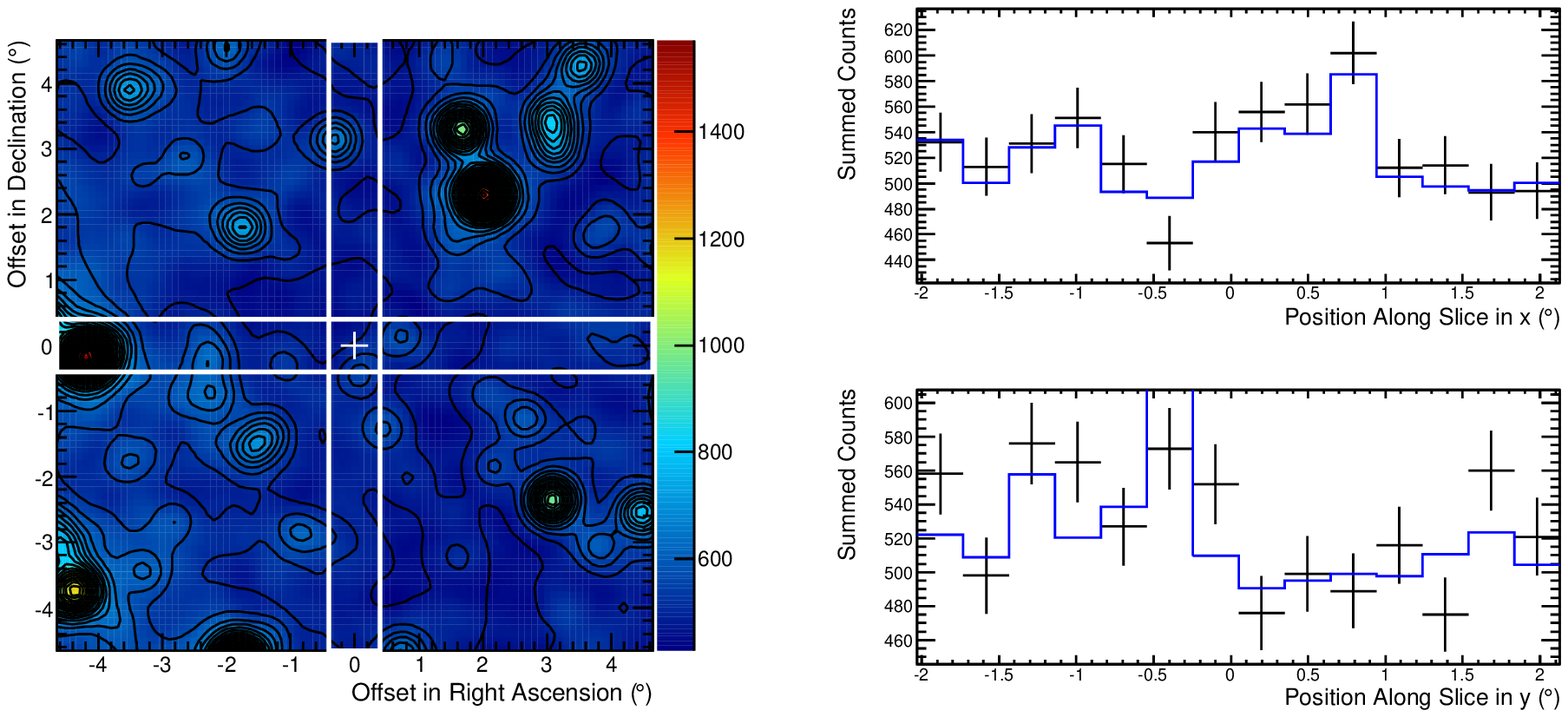}}
    \subfigure{%
     \includegraphics[height=6.8cm]{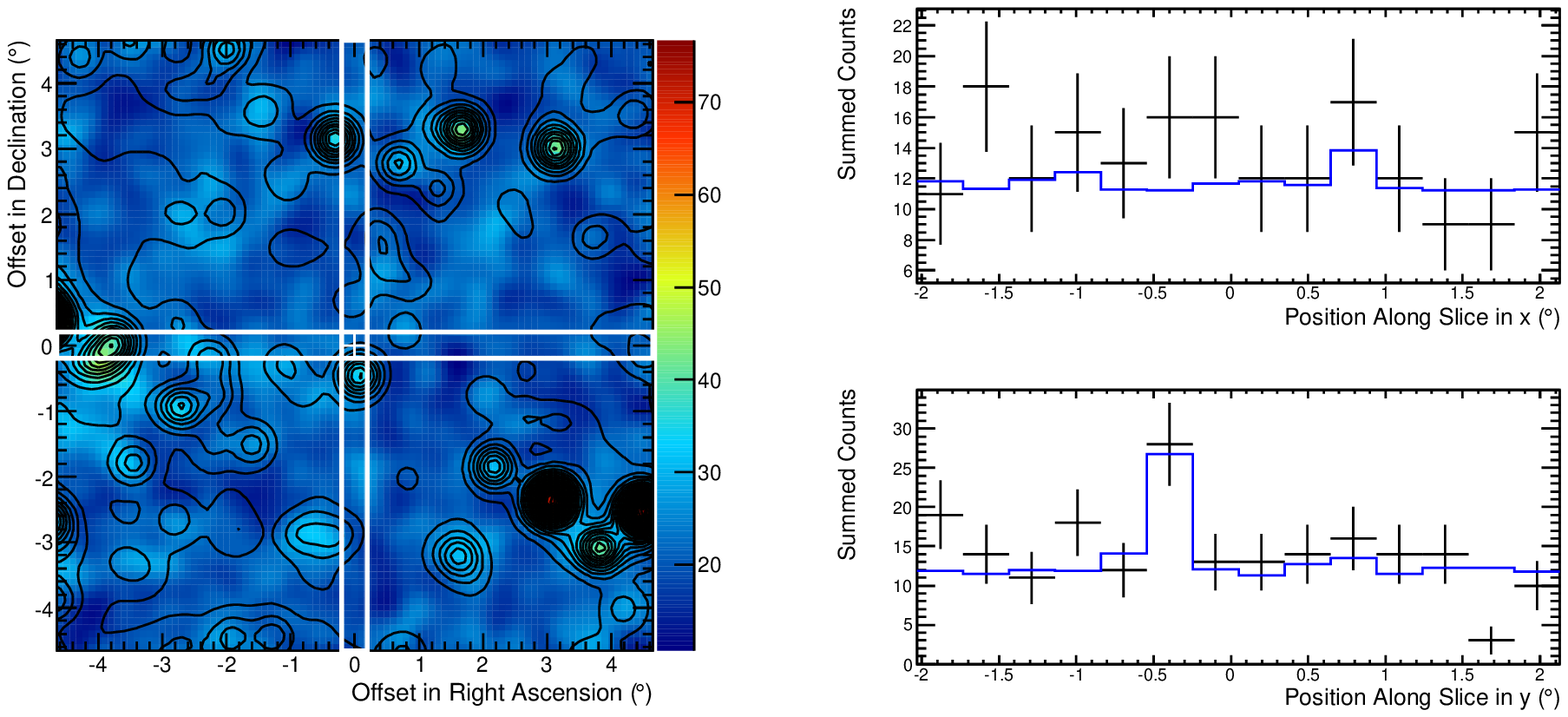}}
  \end{center}
  \caption{Stacked counts maps for the main sample of BCGs above an
    energy ({\bf top}) 300~MeV, ({\bf middle}) 3~GeV and ({\bf
      bottom}) 30~GeV, overlayed with contours representing the
    corresponding stacked model maps (for the null hypothesis). The
    white cross indicates the BCG position in all fields. The stacked
    maps have been smoothed with a Gaussian kernel of rms 0.2\degr. A
    cut in candidate source TS above 16 has been applied, and
    additionally a {\it proximity} cut of 0.2\degr~(to remove fields
    where the candidate position is coincident with a \fermilat
    source). Neglecting also fields for which a 2FGL source brighter
    than $1\times 10^{-8}\,{\rm ph\,cm}^{-2}{\rm s}^{-1}$ is found to
    be within 1\degr~of the central position (see
    \S\ref{ssec:stacking}), these stacks then comprise 109 ({\bf top})
    and 111 ({\bf middle; bottom}) individual fields. The boxed
    regions indicate the area of the slices (to the right of each
    stack) through both maps in the horizontal and vertical
    directions, allowing a 1-D comparison between the data (black
    points) and null model (blue line).  }
  \label{fig:2}
\end{figure*}

\subsection{Sources of Note}
\label{ssec:sources}

\subsubsection{2FGL Sources}

The \fermi source 2FGL\,J0627.1$-$3528 is associated with the BCG in
A\,3392, which is classified as a BL~Lac object. It is listed with a
24-month detection significance of 12.6, $100~{\rm MeV} \leq E \leq
100~{\rm GeV}$ flux of $(1.50\pm 0.18)\times10^{-9}\,{\rm
ph\,cm}^{-2}{\rm s}^{-1}$ and photon index of 1.93 $\pm$ 0.09. With
21 months more data (and in an energy band $300~{\rm MeV} \leq E \leq
300~{\rm GeV}$) we calculate an increased detection significance of
$14.9$, and a consistent (within errors) photon index, as expected.

The \fermi source 2FGL\,J1958.4$-$3012 is associated with the BL~Lac
object RXC\,J1958.2-3011, and may be a BCG (see
\S~\ref{sec:sample}). It is listed with a 24-month detection
significance of 5.1, $100~{\rm MeV} \leq E \leq 100~{\rm GeV}$ flux of
$(5.94 \pm 1.78)\times10^{-10}\,{\rm ph\,cm}^{-2}{\rm s}^{-1}$ and
photon index of 1.9 $\pm$ 0.2. With 21 months more data (and in an
energy band $300~{\rm MeV} \leq E \leq 300~{\rm GeV}$) we calculate an
increased detection significance of $6.8$, and a consistent (within
errors) photon index, as expected.

\subsubsection{Offset Detections: Possible Counterparts} 

TS maps were constructed for candidate sources with TS~$\gtrapprox$16,
to better localise the source position. These can be seen in
Figure~\ref{fig:3} for the BCGs in MACS\,J1347.5$-$1145 (TS $= 43.4$),
RXC\,J0132.6$-$0804 (TS $= 24.0$) and IC\,4991 (TS $= 16.0$), and
illustrate the effect of neglecting to add the candidate to the model:
significant sources are revealed and localised where the baseline
model (containing all local 2FGL sources, and the diffuse Galactic and
extragalactic backgrounds, see \S\ref{sec:analysis}) fails to describe
the \gam-ray emission. These peaks in TS were fitted with parabolic
ellipsoids and the best-fit peak positions are illustrated in the
figure. The contours represent the 95\% confidence regions for the
positions of the sources.

\defcitealias{paggi12}{ATel \#4086} 	
\defcitealias{torresi11}{ATel \#3788} 

Adding a candidate at the position of the BCG in MACS\,J1347.5$-$1145
(in the energy band $300~{\rm MeV} \leq E \leq 300~{\rm GeV}$) yields
a photon index of 2.76 $\pm$ 0.01 and a flux of $(3.81\pm
0.70)\times10^{-9}\,{\rm ph\,cm}^{-2}{\rm s}^{-1}$. The significant
source in the field however, is clearly offset from the BCG position
(the fitted peak in TS is located at Right Ascension (RA) =
207.33\degr, Declination (Dec.) = $-11.55$\degr) and so unlikely to be
associated. Consulting the literature reveals the \fermilat detection
of a GeV flare consistent (within the quoted statistical errors) with
this location in November 2011 \citepalias{torresi11}, and the
flat-spectrum $z=0.34$ Quasi Stellar Object (QSO) PKS\,1346$-$112
\citep{jackson02}, is given as a possible counterpart. Following this
detection, a possible Wide-field Infrared Survey Explorer (WISE)
blazar counterpart was identified, coincident with the radio source
\citepalias{paggi12}. Re-running the likelihood analysis with this new
source included in the model yields a negligible TS at the BCG
position, and a TS of 179.6 (corresponding to a detection significance
of $\approx$13.4$\sigma$) at the peak position, with a corresponding
photon index of 2.44 $\pm$ 0.03 and flux of
$(3.04\pm0.15)\times10^{-8}{\rm ph\,cm}^{-2}{\rm s}^{-1}$. Despite the
high TS value, the transient nature of the source may account for its
absence from the 2FGL.

Adding a candidate at the position of the BCG in RXC\,J0132.6$-$0804
(in the energy band $300~{\rm MeV} \leq E \leq 300~{\rm GeV}$) yields
for the source a photon index of 2.20 $\pm$ 0.02 and a flux of
$(1.42\pm 0.50)\times10^{-9}\,{\rm ph\,cm}^{-2}{\rm s}^{-1}$. There
appear to be two significant sources in the field, the more central of
which (the fitted TS peak is located at RA = 23.25\degr, Dec. =
$-7.98$\degr), is 8\arcmin~from the BCG and may be associated with a
member of the parent cluster. The most significant peak is fitted to a
position at RA = 21.78\degr, Dec. = $-8.37$\degr. The BL~Lac
FBQS\,J0127$-$0821 at a redshift of 0.36 is a plausible \gam-ray
counterpart \citep{plotkin08}. A re-analysis of the field using a
model containing these two localised sources nullifies the BCG TS, and
results in a TS of 28.0 for the closer peak, and 65.0 for the
remaining one, corresponding to $\approx$5.3$\sigma$ and
$\approx$8$\sigma$ detections, respectively. A photon index of 2.36
$\pm$ 0.03 and flux of $(1.14\pm 0.06)\times10^{-8}\,{\rm
ph\,cm}^{-2}{\rm s}^{-1}$ is calculated for the latter, and for the
former a photon index of 1.98 $\pm$ 0.03, and a flux of $(3.06\pm
0.24)\times10^{-9}\,{\rm ph\,cm}^{-2}{\rm s}^{-1}$.

Adding a candidate at the position of the BCG in IC\,4991 (in the
energy band $300~{\rm MeV} \leq E \leq 300~{\rm GeV}$) yields a
photon index of 2.379 $\pm$ 0.009 and a flux of $(1.64\pm
0.62)\times10^{-9}\,{\rm ph\,cm}^{-2}{\rm s}^{-1}$. There appear to
be two potential non-2FGL sources of \gam~radiation in the field.  For
the first, the BCG lies just outside the 95\% confidence contour (the
fitted peak in TS is located at RA = 304.30\degr, Dec. = $-41.22$\degr).
If this TS peak represents a genuine \gam-ray source, it may be
associated with the candidate, or perhaps with the ROSAT source
1RXS\,201731.2$-$411452, a BL~Lac object 6\arcmin~removed
\citep{kollatschny08}. The second TS peak (fit position:
RA = 307.27\degr, Dec. = $-42.72$\degr) may indicate \gam-ray emission
associated with an 81~mJy SUMSS source 5\arcmin~removed, that is not
clearly identified (see \citet{mauch03}). Remodelling the data to
account for these two sources renders the candidate BCG emission
insignificant (reducing the TS to 2.7), and results in a TS of 13.9
for the closer peak, and 41.1 for the peak revealed toward the edge of
the field, roughly corresponding to statistical significances of
3.7$\sigma$ and 6.4$\sigma$, respectively. A photon index of 2.2
$\pm$ 0.2 and flux of $(8.58\pm 3.15)\times10^{-9}\,{\rm
ph\,cm}^{-2}{\rm s}^{-1}$ is calculated for the significant source.

\begin{figure}    
  \begin{center}
    \subfigure{%
      \includegraphics[height=7cm]{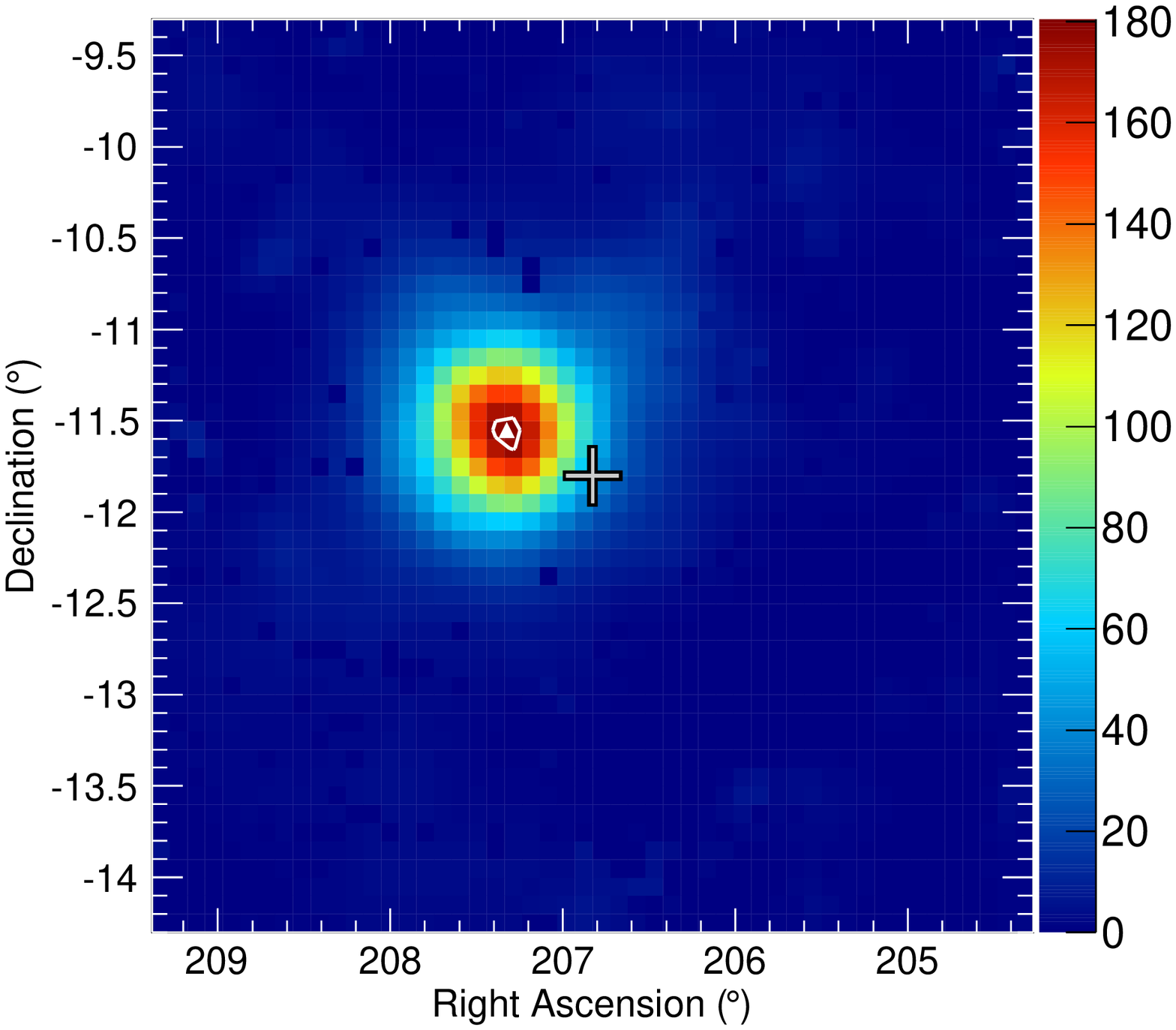}}
    \subfigure{%
     \includegraphics[height=7cm]{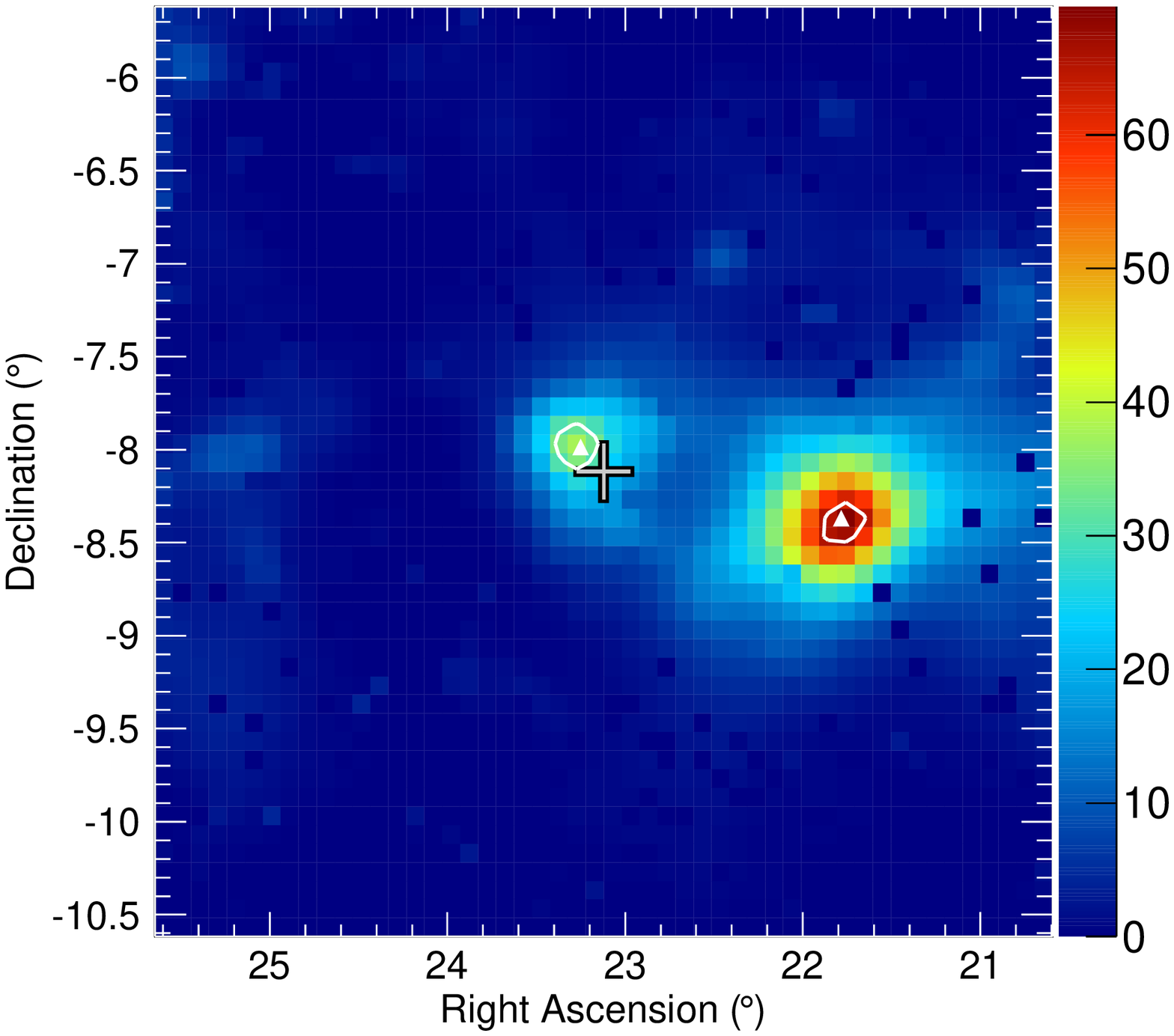}}
    \subfigure{%
     \includegraphics[height=7cm]{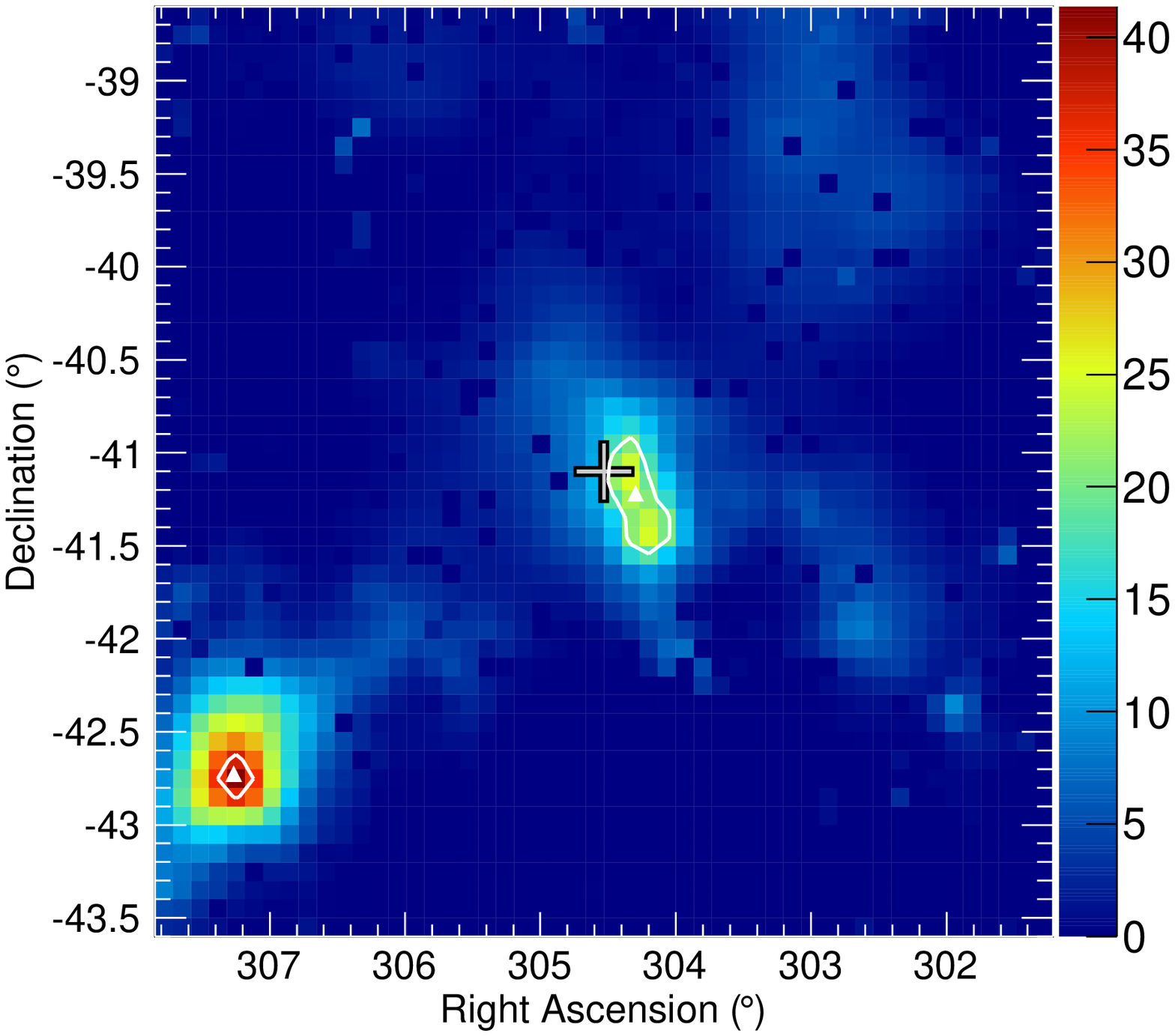}}
  \end{center}
  \caption{TS map for candidate sources in MACS\,J1347.5$-$1145 ({\bf
      top}), RXC\,J0132.6$-$0804 ({\bf middle}), and IC\,4991 ({\bf
      bottom}) produced from $300~{\rm MeV} \leq E \leq 300~{\rm GeV}$
      data. The BCG position is indicated by the central black/grey
      cross. Peaks in TS have been fitted with parabolic ellipsoids,
      the fit position for which is shown by a white triangle and
      given in table~\ref{tab:sources}. The 95\% confidence contours
      on source locations are shown in white.}
  \label{fig:3}
\end{figure}

Table~\ref{tab:sources} summarises the detections confirmed after a
remodelling to account for sources of \gam-ray emission in the three
fields described above was carried out. The TS maps were re-calculated
based on the new output models, which were found to describe the
emission adequately.

\begin{table}
    \begin{center}
    \caption{A summary of non-2FGL \fermi detections of sources in fields
    within our sample, giving the fitted position (and error) of the
    TS peak, and the output TS after remodelling the data. No attempt
    is made to account for trials factors due to the large sample
    considered, but we note that a TS value of $>$25 still corresponds to
  a post-trials significance of $>$4$\sigma$.}\label{tab:sources}
    \begin{tabular}{llllr}
      \hline
      \hline
      Name                                     & TS{$^{\ast}$}    & RA{$^{\dag}$}~(\degr) & Dec{$^{\ddag}$}~(\degr)     & Error~(\degr) \\
      \hline
      FERMI J1350$-$1140{$^{\dag\dag}$}  & 179.8                  & 207.33                       & $-$11.55                         & 0.11 \\
      FGL\,J0133.0$-$0758                      & 28.2                   & 23.25                        & $-$7.98                          & 0.16 \\
      FGL\,J0127.1$-$0822                      & 64.8                   & 21.78                        & $-$8.37                          & 0.14 \\
      FGL\,J2029.0$-$4243                      & 41.3                   & 307.27                       & $-$42.72                         & 0.22 \\
      \hline
    \end{tabular}
 \end{center}
\vspace{0.5mm} 
  \scriptsize{
    $^{\ast}$ Test Statistic\\
    $^{\dag}$ Right Ascension\\
    $^{\ddag}$ Declination\\
    $^{\dag\dag}$ Detected by the Fermi Large Area Telescope Collaboration in November 2011 (ATel \#3788)
  } 
\end{table}

\subsubsection{$\gamma$-ray Emission from A\,2055?}

Using the $3\,{\rm GeV} \leq E \leq 300\,{\rm GeV}$ data, analysis of
the BCG in A\,2055 (which is classified as a BL~Lac object) results in
a TS of 15.2. The corresponding TS map is shown in
Figure~\ref{fig:4}. A hint of emission from the position of the BCG
can be clearly seen. The position of the BCG is contained within the
95\% confidence region for the source location. With this candidate
alone added, the model sufficiently describes the \gam-ray emission
within the field. The source is not detected in the energy band
$30\,{\rm GeV} \leq E \leq 300\,{\rm GeV}$, and the TS drops to 8.0
using the larger $300\,{\rm MeV} \leq E \leq 300\,{\rm GeV}$ dataset.

\begin{figure}
  \centering
  \includegraphics[height=7cm]{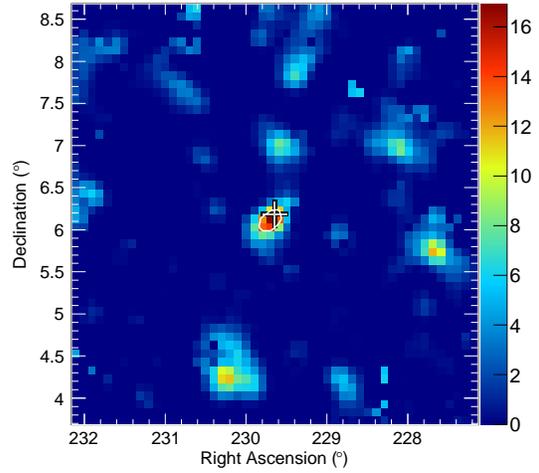}
  \caption{TS map for the candidate source in A\,2055, produced from
    $3~{\rm GeV} \leq E \leq 300~{\rm GeV}$ data. The BCG position is
    indicated by the central black/grey cross. The 95\% confidence
    contour on the \gam-ray source position is shown in white.}
  \label{fig:4}
\end{figure}

\section[]{Discussion}
\label{sec:discussion}

There is no statistically significant evidence for emission within the
stacked sample. The upper limits derived on an average source within
the class of BCGs represent at least an order of magnitude improvement
on the individual upper limits presented in
Table~\ref{tab:sample}. The derived limits are consistent with the
expectation of a $\sim1/\sqrt{N_{\rm obj}}$ scaling for the
background-limited lower energy bands and $\sim1/{N_{\rm obj}}$ for
highest energies where the \fermilat becomes signal limited even in
the stacked sample, where $N_{\rm obj}$ is the number of fields
stacked.

To derive a limit on the \gam-ray to radio flux ratio ($F_{1 \rm
GeV}/F_{1.4 \rm GHz}$) for the core emission of the sample we consider
the {\it core} radio flux (some fraction of the emission given in
Table~\ref{tab:sample}) of each BCG \citep{hogan}, and find an average
radio flux of $2.37 \times 10^{-15}{\rm erg}\,{\rm cm}^{-2}\,{\rm
s}^{-1}$ and an average luminosity of $1.8 \times 10^{41} {\rm
erg\,s}^{-1}$ (at an average redshift of 0.13). Given the upper limit
on the stacked emission, the average \gam-ray flux evaluated at 1~GeV
and assuming a photon index of 2, is less than $3.55 \times
10^{-14}\,{\rm erg}\,{\rm cm}^{-2}\,{\rm s}^{-1}$, and the average
\gam-ray luminosity is less than $1.35 \times 10^{42} {\rm
erg\,s}^{-1}$. The average $F_{1 \rm GeV}/F_{1.4 \rm GHz}$ of the
sample is therefore $<$15. The \gam-ray flux-limit for the sample
corresponds to only 0.1\% of the flux of NGC\,1275, which has $F_{1
\rm GeV}/F_{1.4 \rm GHz}\approx$120 ($\sim90$ excluding \gam-ray
flares). The average core radio flux is $\approx$1\% of that of
NGC\,1275. The lack of a $\gamma$-ray signal in the stack therefore
implies that the average core-radio selected BCGs have spectral energy
distributions with a different shape to that of NGC\,1275, with
significantly lower \gam-ray flux relative to core radio flux, though
the current highly variable state of \gam-ray emission in this active
galaxy \citep{kataoka10} may account for the difference in $F_{1 \rm
GeV}/F_{1.4 \rm GHz}$.

NGC\,1275 and M\,87 are the closest cooling-core BCGs, so the detected
$\gamma$-ray emission is unlikely to be beamed toward us if they are
drawn from a random distribution of sight lines. However, a role for
beaming in NGC\,1275 cannot be excluded and may play a role in its
unusual \gam-ray brightness (see above). It has been postulated (see
\citet{falceto10a}) that the central AGN comprises a viscous accretion
disk tilted with respect to the rotating super-massive black hole,
bringing about precessing jets as a result of torques acting on the
disk. Very Long Baseline Interferometry (VLBI) maps of the
radio source can place constraints on the inclination of the system:
to be satisfied, some curvature of the jet away from the line of sight
is implied \citep{dunn06}. This said, for those BCGs with bona fide
BL~Lac nuclei, for example A\,2055 and A\,3392, which are
significantly more distant and comparable to most of the parent
sample, the probability of observing a few sources close to the jet
direction is significant.

We can assess the number of local clusters hosting a \fermilat AGN by
cross-correlating the Abell cluster catalogue of 5250 clusters
(including the supplementary systems) \citep{abell89}, with the
2FGL. Of the 85 coincidences within 20$'$, only 11 are consistent with
the 2FGL source being a cluster member and of those only four being
the BCG (NGC\,1275 in A\,426/Perseus, PKS\,0625$-$35 in A\,3392,
PKS\,2035$-$714 in A\,3701 and PKS\,2316$-$423 in S1111). The latter
two clusters are not detected above the X-ray flux limit for the eBCS
or REFLEX catalogues, so were not considered in the analysis
above. Therefore, the chance of finding a \fermi source in any
particular cluster is relatively small but is significant.

The added complication with the analysis is that the \fermilat spatial
resolution and source density ($\approx$0.05 per square degree)
implies chance coincidences on the order of a few when more than 100
positions are considered. This is illustrated in Figure~\ref{fig:3},
given that the TS output from the initial binned likelihood analysis
of the field was in each case not exclusively associated with the BCG
under scrutiny, but a nearby source of emission not described by the
model. The angular resolution of the \fermilat is sufficient
(particularly above 1~GeV) to isolate nearby background sources from
the position of interest in most cases, but the likelihood analysis is
not robust enough to accurately disentangle the \gam~emissions in such
an event.

Four {\it detections} (that is, of TS$>$25 emission) have been made of
sources not contained in the 2FGL (see Table~\ref{tab:sources}). For
those above a TS of $\sim$50, their absence from the 2FGL implies
temporal variability in the emission, as appears to be the case for
the source FERMI J1350$-$1140. A more detailed study of these sources
is left to future work. The tentative evidence for emission from the
BCG in A\,2055 will be refuted or validated in time: if the
significance of emission from the source position continues to
increase, it will be the fourth such source to be detected using {\it
  Fermi}; the second BCG hosting a BL~Lac.

\section[]{Concluding Remarks}
\label{sec:conclusion}

The \fermilat detection of a few sources in clusters of galaxies sets
important limits on the energetics and emission mechanisms at work in
the cores of the most massive member galaxies. The detections of BCGs
using \fermi are strongly biased toward the most radio bright systems
or those that are dominated by a beamed core. Our statistical upper
limit from a larger sample of sources with lower radio flux indicates
that the average core-radio selected BCG is more than an order of
magnitude less $\gamma$-ray luminous than, for example, NGC\,1275.
Furthermore the ratio of \gam-ray to radio flux is $<$15, compared
with $F_{1 \rm GeV}/F_{1.4 \rm GHz}\approx$120 for NGC\,1275. The BCG
in A\,2055, which hosts a BL~Lac, may soon be detectable in HE
\gam~rays (given a longer exposure using \fermi), and if so, will
corroborate this conclusion.

Detections of a number of new sources are declared; though none of
these appear to be unambiguously associated with the BCG, the
slightly-offset emission in the field of RXC\,J0132.6$-$0804 might be
associated with the candidate, or else another cluster member galaxy,
and in the field of IC\,4991 may (if genuine) in part originate from
the BCG. This work also illustrates then, the need to carefully assess
all the possible sources of emission that could explain the \fermilat
detections found in clusters before attributing them to (for example)
cluster merger-related shocks or DM annihilation.

\section*{Acknowledgments}

The authors would like to thank Stefan Ohm for useful comments and
discussions. This work has made use of public \fermi data and Science
Tools provided by the \fermi Science Support Centre. K. L.  Dutson and
R. J. White acknowledge support from the STFC and J. A. Hinton from
the Leverhulme Trust. 

\bibliography{Dutson2012_astro-ph}
\bibliographystyle{mn2e}

\appendix
\bsp
\label{lastpage}

\end{twocolumn}
\end{document}

%% file: table.tex
\renewcommand{\thefootnote}{\fnsymbol{footnote}}
\begin{center}
\begin{longtable}{llllrrr}
\caption{A summary of the 114 brightest cluster galaxies of the main sample, and their candidate {\it Fermi} source properties, including the upper limit on the $\gamma$-ray flux above 300~MeV. The radio emission is classified as core-dominated (`c'), extended-with-core-contribution (`e') or BL~Lac-associated (`b'). \label{tab:sample}}\\
\hline\hline

\T Parent Cluster & {\it l}{$^{\ast}$} (\degr)& {\it b}{$^{\dag}$} (\degr)& {\it z} & 1.4~GHz{$^{\ddag}$} Radio & TS{$^{\dag\dag}$} & $\gamma$-ray Flux Upper~~ \\
 & & & & Flux (mJy)~~~ & & Limit (${\rm ph}\,{\rm cm}^{-2}\,{\rm s}^{-1}$)\\
\hline

\endfirsthead

 \hline
\multicolumn{7}{c}%
{\T \tablename\ \thetable{} -- {\it continued}}\\
\hline
\T Parent Cluster & {\it l}{$^{\ast}$}(\degr) & {\it b}{$^{\dag}$}(\degr) & {\it z} & 1.4~GHz{$^{\ddag}$} Radio & TS{$^{\dag\dag}$} & $\gamma$-ray Flux Upper~~ \\
 & & & & Flux (mJy)~~~ & & Limit (${\rm ph}\,{\rm cm}^{-2}\,{\rm s}^{-1}$)\\
\hline
\endhead

\hline
\endfoot

\hline \hline
\endlastfoot
RXCJ0000.1+0816 &   101.9 &   -52.5 & 0.04 & 84 c & 0 &  \T $   2.2 \cdot 10^{-10}$ \\
A85 &   115.3 &   -72.0 & 0.056 & 58 c & 0 & $   2.8 \cdot 10^{-10}$ \\
Z235 &   120.8 &   -38.4 & 0.083 & 50 c & 3.2 & $   5.7 \cdot 10^{-10}$ \\
3C31 &   126.9 &   -30.3 & 0.016 & 1108 e & 2.2 & $   5.8 \cdot 10^{-10}$ \\
RXCJ0132.6-0804 &   152.0 &   -68.6 & 0.15 & 308 c & 24.0 & $   1.6 \cdot 10^{-9}$ \\
MACSJ0242.5-2132 &   206.6 &   -64.1 & 0.31 & 1255 c & 0 & $   1.9 \cdot 10^{-10}$ \\
A3112 &   253.0 &   -56.1 & 0.075 & {\it 1915} c & 11.2 & $   8.2 \cdot 10^{-10}$ \\
A496 &   209.7 &   -36.5 & 0.033 & 121 c & 5.8 & $   9.2 \cdot 10^{-10}$ \\
RXCJ0439.0+0520 &   191.4 &   -26.2 & 0.21 & 82 c & 6.0 & $   8.2 \cdot 10^{-10}$ \\
NGC1650 &   214.0 &   -34.9 & 0.036 & 91 c & 0 & $   5.0 \cdot 10^{-10}$ \\
S555 &   243.6 &   -26.3 & 0.044 & 319 c & 0 & $   2.1 \cdot 10^{-10}$ \\
A3378 &   241.9 &   -24.0 & 0.14 & 1337 c & 0 & $   2.9 \cdot 10^{-10}$ \\
A3392 &   243.5 &   -20.0 & 0.055 & 4513 b & 221.0 & --~~~~~~~\\
PKS0745-191 &   236.5 &     3.0 & 0.1 & 2373 c & 13.7 & $   8.2 \cdot 10^{-10}$ \\
A646 &   172.7 &    34.6 & 0.13 & 54 c & 0 & $   8.3 \cdot 10^{-11}$ \\
4C+55.16 &   162.3 &    36.6 & 0.24 & 8284 c & 1.6 & $   4.9 \cdot 10^{-10}$ \\
Hydra-A &   243.0 &    25.1 & 0.054 & 40850 c & 2.7 & $   4.8 \cdot 10^{-10}$ \\
MACSJ1133.2+5008 &   150.7 &    62.6 & 0.39 & 846 e & 0 & $   2.3 \cdot 10^{-10}$ \\
A1348 &   277.4 &    47.1 & 0.12 & 158 c & 0 & $   1.4 \cdot 10^{-10}$ \\
A1366 &   132.6 &    48.5 & 0.12 & 237 b & 0 & $   3.0 \cdot 10^{-10}$ \\
NGC4696 &   302.5 &    21.6 & 0.0098 & {\it 5674} e & 0 & $   3.4 \cdot 10^{-10}$ \\
MACSJ1347.5-1145 &   324.1 &    48.8 & 0.45 & 48 c & 43.4 & --~~~~~~~\\
A1795 &    33.9 &    77.2 & 0.062 & 925 e & 0 & $   1.9 \cdot 10^{-10}$ \\
RXCJ1350.3+0940 &   344.3 &    67.7 & 0.13 & 293 c & 0 & $   1.0 \cdot 10^{-10}$ \\
A3581 &   323.2 &    32.9 & 0.023 & 646 c & 7.4 & $   1.1 \cdot 10^{-9}$ \\
A1885 &    83.2 &    66.6 & 0.089 & 49 c & 0 & $   1.2 \cdot 10^{-10}$ \\
S780 &   341.0 &    35.1 & 0.24 & 106 c & 1.5 & $   3.1 \cdot 10^{-10}$ \\
RXCJ1504.1-0248 &   355.1 &    46.2 & 0.22 & 62 c & 7.3 & $   9.6 \cdot 10^{-10}$ \\
A2052 &     9.5 &    50.1 & 0.035 & 5500 c & 0 & $   2.4 \cdot 10^{-10}$ \\
A2055 &     8.9 &    49.3 & 0.1 & 498 b & 8.0 & $   8.4 \cdot 10^{-10}$ \\
RXCJ1524.2-3154 &   337.1 &    20.7 & 0.1 & 50 c & 2.0 & $   5.3 \cdot 10^{-10}$ \\
RXCJ1558.3-1410 &   356.6 &    28.7 & 0.097 & 461 c & 1.7 & $   6.8 \cdot 10^{-10}$ \\
A2199 &    63.0 &    43.7 & 0.03 & 3681 e & 1.5 & $   5.9 \cdot 10^{-10}$ \\
NGC6338 &    85.9 &    35.4 & 0.028 & 57 c & 0 & $   3.5 \cdot 10^{-10}$ \\
Z8193 &    67.6 &    34.7 & 0.18 & 134 c & 0 & $   2.7 \cdot 10^{-10}$ \\
A2270 &    83.1 &    33.8 & 0.24 & 144 c & 2.6 & $   5.1 \cdot 10^{-10}$ \\
Z8276 &    58.0 &    27.6 & 0.076 & 92 c & 0 & $   5.6 \cdot 10^{-10}$ \\
RXCJ1750.2+3504 &    60.6 &    27.0 & 0.17 & 69 c & 0 & $   1.5 \cdot 10^{-10}$ \\
E1821+643 &    94.1 &    27.4 & 0.3 & 94 e & 1.3 & $   2.7 \cdot 10^{-10}$ \\
RXCJ1832.5+6848 &    99.2 &    26.8 & 0.2 & 150 c & 0 & $   1.1 \cdot 10^{-10}$ \\
A3639 &   347.0 &   -26.3 & 0.15 & {\it 139} c & 0 & $   2.1 \cdot 10^{-10}$ \\
RXCJ1931.6-3354 &     5.2 &   -22.5 & 0.097 & 886 c & 0 & $   2.8 \cdot 10^{-10}$ \\
MACSJ1931.6-2634 &    12.6 &   -20.1 & 0.35 & 223 c & 7.4 & $   1.0 \cdot 10^{-9}$ \\
RXCJ1958.2-3011 &    11.0 &   -26.8 & 0.12 & 128 b & 46.3 & --~~~~~~~\\
Cyg-A &    76.3 &     5.8 & 0.056 & 1590000 e & 4.5 & $   7.9 \cdot 10^{-10}$ \\
S851 &   351.0 &   -32.6 & 0.0097 & {\it 139} c & 0 & $   9.3 \cdot 10^{-10}$ \\
RXCJ2014.8-2430 &    18.4 &   -28.5 & 0.16 & 230 e & 0 & $   3.4 \cdot 10^{-10}$ \\
A2390 &    74.0 &   -27.8 & 0.23 & 236 c & 11.2 & $   4.3 \cdot 10^{-10}$ \\
RXCJ2213.1-2754 &    22.3 &   -55.0 & 0.061 & 76 c & 0 & $   2.8 \cdot 10^{-10}$ \\
A3880 &    18.1 &   -58.5 & 0.058 & 228 c & 4.0 & $   5.7 \cdot 10^{-10}$ \\
A2597 &    65.4 &   -64.9 & 0.085 & 1875 e & 0 & $   1.9 \cdot 10^{-10}$ \\
A2627 &   101.8 &   -35.9 & 0.13 & 434 b & 9.8 & $   8.3 \cdot 10^{-10}$ \\
A2634 &   103.6 &   -33.1 & 0.03 & 1037 e & 0 & $   1.9 \cdot 10^{-10}$ \\
RXCJ0137.2-0911 &   156.3 &   -69.1 & 0.041 & 178 e & 0 & $   3.1 \cdot 10^{-10}$ \\
A262 &   136.6 &   -25.1 & 0.017 & 67 e & 0 &  \T $   4.3 \cdot 10^{-10}$ \\
A2984 &   256.6 &   -68.9 & 0.1 & 232 c & 0 & $   1.5 \cdot 10^{-10}$ \\
A3017 &   256.6 &   -65.7 & 0.22 & {\it 127} c & 3.3 & $   5.1 \cdot 10^{-10}$ \\
Z808 &   175.4 &   -47.3 & 0.17 & 393 e & 0 & $   9.6 \cdot 10^{-11}$ \\
A407 &   150.7 &   -19.9 & 0.048 & 661 c & 0 & $   1.7 \cdot 10^{-10}$ \\
RXCJ0331.1-2100 &   212.3 &   -53.2 & 0.19 & 168 c & 0 & $   2.5 \cdot 10^{-10}$ \\
NGC1399 &   236.8 &   -53.6 & 0.0051 & 2500 e & 2.9 & $   4.8 \cdot 10^{-10}$ \\
A3165 &   226.4 &   -51.4 & 0.14 & 921 c & 0 & $   1.4 \cdot 10^{-10}$ \\
RXCJ0359.1-0319 &   193.4 &   -39.3 & 0.12 & 188 e & 0 & $   2.8 \cdot 10^{-10}$ \\
RXCJ0425.8-0833 &   203.4 &   -36.2 & 0.04 & 112 e & 0 & $   2.5 \cdot 10^{-10}$ \\
MACSJ0429.6-0253 &   197.9 &   -32.6 & 0.4 & 139 c & 0 & $   1.3 \cdot 10^{-10}$ \\
A499 &   218.5 &   -38.3 & 0.15 & 158 e & 0 & $   4.8 \cdot 10^{-10}$ \\
RXCJ0501.4+0110 &   198.5 &   -23.7 & 0.12 & 121 c & 0 & $   3.2 \cdot 10^{-10}$ \\
MACSJ0520.7-1328 &   215.3 &   -26.1 & 0.34 & 93 c & 0 & $   2.8 \cdot 10^{-10}$ \\
A3360{$^{\ast \ast}$} &   249.4 &   -30.7 & 0.085 & {\it 222} c & 0 & $   2.2 \cdot 10^{-10}$ \\
S547 &   254.2 &   -30.0 & 0.051 & {\it 143} e & 0 & $   3.1 \cdot 10^{-10}$ \\
A3380 &   257.2 &   -27.3 & 0.055 & {\it 1455} e & 1.0 & $   5.1 \cdot 10^{-10}$ \\
A3396 &   250.0 &   -21.6 & 0.18 & {\it 614} e & 2.5 & $   2.4 \cdot 10^{-10}$ \\
A795 &   217.1 &    40.2 & 0.14 & 116 c & 1.1 & $   1.8 \cdot 10^{-10}$ \\
S617 &   253.3 &    23.3 & 0.034 & 105 e & 7.3 & $   7.5 \cdot 10^{-10}$ \\
A907 &   249.4 &    33.3 & 0.17 & 69 c & 0 & $   1.5 \cdot 10^{-10}$ \\
A923 &   205.3 &    53.3 & 0.12 & 73 c & 0 & $   1.9 \cdot 10^{-10}$ \\
Z3179 &   228.7 &    53.1 & 0.14 & 94 c & 6.6 & $   7.4 \cdot 10^{-10}$ \\
Z3916 &   144.7 &    54.4 & 0.21 & 66 c & 0 & $   9.0 \cdot 10^{-11}$ \\
A1361 &   153.4 &    66.6 & 0.12 & 864 e & 0 & $   3.0 \cdot 10^{-10}$ \\
A3490 &   287.8 &    26.5 & 0.07 & 368 e & 0 & $   2.0 \cdot 10^{-10}$ \\
A1412 &   128.4 &    43.1 & 0.083 & 73 e & 0 & $   2.0 \cdot 10^{-10}$ \\
MACSJ1206.2-0847 &   284.5 &    52.4 & 0.44 & 161 e & 0 & $   1.1 \cdot 10^{-10}$ \\
A1644 &   305.0 &    45.4 & 0.047 & 98 c & 0 & $   5.8 \cdot 10^{-10}$ \\
A1677 &    84.0 &    85.1 & 0.18 & 76 c & 0 & $   1.7 \cdot 10^{-10}$ \\
NGC5098 &    78.8 &    81.3 & 0.036 & 83 c & 0 & $   3.7 \cdot 10^{-10}$ \\
A3565 &   313.6 &    28.0 & 0.012 & 3000 e & 0 & $   5.4 \cdot 10^{-10}$ \\
MACSJ1411.3+5212 &    97.6 &    60.8 & 0.46 & 22720 e & 0 & $   3.9 \cdot 10^{-10}$ \\
A2204 &    21.2 &    33.2 & 0.15 & 70 c & 0 & $   1.3 \cdot 10^{-10}$ \\
HerA &    23.1 &    28.9 & 0.15 & 34000 e & 3.6 & $   3.3 \cdot 10^{-10}$ \\
NGC6269 &    49.1 &    35.9 & 0.035 & 51 c & 3.0 & $   2.5 \cdot 10^{-10}$ \\
RXCJ1715.1+0309 &    24.5 &    22.8 & 0.13 & 165 c & 0 & $   2.7 \cdot 10^{-10}$ \\
RXCJ1720.1+2637 &    49.3 &    30.9 & 0.16 & 89 e & 0 & $   1.4 \cdot 10^{-10}$ \\
RXCJ1840.6-7709 &   317.3 &   -25.8 & 0.019 & {\it 1152} e & 0 & $   2.3 \cdot 10^{-10}$ \\
RXCJ1852.1+5711 &    87.1 &    22.4 & 0.11 & 51 c & 0 & $   4.5 \cdot 10^{-10}$ \\
A2312 &    99.0 &    24.8 & 0.095 & 79 c & 5.4 & $   6.3 \cdot 10^{-10}$ \\
A3638 &   355.4 &   -24.0 & 0.077 & {\it 258} c & 0 & $   2.4 \cdot 10^{-10}$ \\
IC4991 &   359.9 &   -33.2 & 0.019 & {\it 109} c & 16.0 & $   1.6 \cdot 10^{-9}$ \\
S868 &    22.9 &   -29.1 & 0.056 & 229 c & 0 & $   2.1 \cdot 10^{-10}$ \\
RXCJ2034.9-2143 &    23.2 &   -32.0 & 0.19 & 82 c & 0 & $   3.4 \cdot 10^{-10}$ \\
RXCJ2043.2-2144 &    23.9 &   -33.8 & 0.2 & 295 e & 0 & $   1.8 \cdot 10^{-10}$ \\
A2331 &    40.9 &   -31.7 & 0.079 & 106 e & 5.3 & $   3.5 \cdot 10^{-10}$ \\
RXCJ2101.5-1317 &    35.4 &   -34.8 & 0.028 & 1200 e & 0 & $   1.9 \cdot 10^{-10}$ \\
RBS1712 &    22.2 &   -38.8 & 0.19 & 67 c & 0 & $   1.9 \cdot 10^{-10}$ \\
APM699 &   338.1 &   -45.7 & 0.082 & {\it 152} e & 11.7 & $   9.8 \cdot 10^{-10}$ \\
RXCJ2151.3-5521 &   339.2 &   -47.1 & 0.038 & {\it 108} e & 9.3 & $   3.0 \cdot 10^{-10}$ \\
A2384 &    33.4 &   -48.4 & 0.096 & 54 c & 0 & $   3.1 \cdot 10^{-10}$ \\
A2389 &    53.7 &   -41.8 & 0.15 & 78 c & 0 & $   3.6 \cdot 10^{-10}$ \\
A2415 &    54.0 &   -45.1 & 0.058 & 187 c & 0 & $   1.4 \cdot 10^{-10}$ \\
A3856 &     2.8 &   -56.2 & 0.14 & 186 e & 0 & $   1.6 \cdot 10^{-10}$ \\
A2445 &    83.0 &   -30.8 & 0.17 & 77 c & 6.0 & $   4.3 \cdot 10^{-10}$ \\
A3911 &   336.7 &   -55.4 & 0.097 & {\it 418} e & 0 & $   1.5 \cdot 10^{-10}$ \\
S1101 &   348.4 &   -64.8 & 0.056 & {\it 472} c & 0 &  \T $   3.0 \cdot 10^{-10}$ \\
A4023 &   304.7 &   -31.7 & 0.19 & {\it 143} e & 0 & $   2.9 \cdot 10^{-10}$ \\
A2665 &    97.0 &   -53.6 & 0.057 & 56 c & 0 & $   1.7 \cdot 10^{-10}$ \\
\footnotetext[1]{Galactic longitude}
\footnotetext[2]{Galactic latitude}
\footnotetext[3]{Taken from the NVSS. If italicised, the values instead denote the 843~MHz radio flux taken from the SUMSS}
\footnotetext[8]{Test Statistic -- for clarity TS values $<$1 are shown as zero}
\footnotetext[7]{Despite a low TS, this source was not included in the stacked analysis (see \S\ref{ssec:stacking}) due to the proximity of a bright 2FGL source, as detailed in the text}
\end{longtable}
\end{center}